\documentclass[12pt,preprint]{aastex}
\usepackage{epsfig}
\usepackage{natbib}

\begin{document}
\voffset-0.5cm
\newcommand{\gsim}{\hbox{\rlap{$^>$}$_\sim$}}
\newcommand{\lsim}{\hbox{\rlap{$^<$}$_\sim$}}

\title {Common origin of the  high energy astronomical \\
         gamma rays, neutrinos and cosmic ray  positrons?}

\author{Shlomo Dado\altaffilmark{1} and Arnon Dar\altaffilmark{1}}

\altaffiltext{1}{Physics Department, Technion, Haifa 32000, Israel}

\begin{abstract} 
We show that the observed fluxes, spectra and sky distributions of the 
high energy astronomical neutrinos,  gamma rays
and cosmic ray positrons satisfy the simple relations expected from their 
common production in hadronic collisions in/near source
of high energy cosmic rays with diffuse matter. 

\end{abstract}


\maketitle

\section{Introduction}

Nearly half a century ago it has already been realized that the decay of 
mesons produced in the interactions of high energy cosmic rays (CRs) with 
matter and radiation in/near the cosmic ray sources, in the interstellar 
medium (ISM), and in the intergalactic medium (IGM), can produce 
detectable fluxes of high energy gamma rays (${\rm \gamma}$'s), neutrinos 
(${\rm \nu}$'s) and positrons (${\rm e^+}$'s), which carry unique 
information on the ultra relativistic universe (e.g., Zatsepin \& Kuzmin 
1966; Berezinsky \& Smirnov~1975; Margolis et al.~1978; Eichler~1978; 
Eichler \& Schramm 1978; Stecker~1979). In particular, interaction of 
cosmic rays with diffuse matter and/or radiation in/near the cosmic ray 
sources, in the ISM of Galaxies and the IGM has long been recognized as 
possible point and diffuse sources of very high energy Galactic and 
extragalactic ${\rm \gamma}$-rays and neutrinos (see, e.g., 
Dar~1983,1985,1991; Halzen et al.~1990; Berezinsky 1991; Stecker et 
al.~1991; Mannheim~1993; Berezinsky et al. 1993,1994; Domokos et al.~1993; 
Dar \& Shaviv 1996; Halzen 1996; Dar \& Laor 1997).

The successful detection of neutrinos from supernova 1987A with the deep
underground Kamiokande and IMB water Cherenkov telescopes, of 
Galactic and extragalactic point sources of very high energy ${\rm 
\gamma}$-rays with ground based atmospheric and water Cherenkov telescopes
(see, e.g., Weekes et al.~1989; Punch et al.~1992), and of
point sources and  diffuse Galactic and extragalactic backgrounds 
of high energy ${\rm \gamma}$-rays with 
the Compton Gamma Ray Observatory (Hunter et al.~1997; Sreekumar et 
al.~1998), have led to the development of  new 
generations of much larger and more sensitive telescopes of
high energy ${\rm \gamma}$-rays, neutrinos   
and cosmic rays, which have already made several
breakthrough observations. Such breakthroughs include the discovery of a 
diffuse high energy neutrino background radiation (NBR) with the megaton 
IceCube detector under the south pole (Aartsen et al.~2013,2014,2015), the 
accurate measurement of the diffuse high energy gamma-ray background 
radiation (GBR) with the Large Area Telescope (LAT) aboard the Fermi 
satellite (Ackermann et al.~2012,2015), and the precise measurements of 
the high energy  local background of cosmic ray positrons with the Payload 
for Antimatter Matter Exploration and Light-nuclei Astrophysics (PAMELA) 
satellite (Adriani et al.~2013) and the Alpha Magnetic Spectrometer (AMS) 
aboard the International Space Station (Aguilar et al.~2014).

Possible sources and the detailed properties of the observed 
astronomical neutrinos, ${\rm \gamma}$-rays and positrons were discussed 
in the discovery papers and in a large number of publications, which 
followed the observations (for a recent review see, e.g., Anchordoqui et 
al.~ 2014a). But, most of the published theoretical estimates of 
the fluxees of the high 
energy neutrinos, ${\rm \gamma}$-rays and positrons from point 
sources and diffuse sources, both before and after these recent 
breakthrough observations, were fraught with uncertainties concerning the 
production mechanisms of the high energy neutrinos, ${\rm \gamma}$-rays 
and positrons and the properties of their sources, 
which have left the origin and observed properties of these radiations 
essentially unsolved puzzles.

In view of such dificulties, several authors have tried alternative 
approaches towards resolving these puzzles, such as deriving upper bounds 
on the observed fluxes of the high energy astronomical neutrinos, 
gamma-rays and positrons, or general relations between them, which may 
indicate their origin. E.g., a highly cited "model-independent upper 
bound" on the energy flux of extragalactic NBR above 100 TeV was derived 
by Waxman \& Bahcall (1999). It was based on the assumption that the NBR 
is produced by the decay of ${\rm \pi^+}$'s produced by the extragalactic 
cosmic ray protons in proton-photon collisions in the extragalctic CR 
sources. However, as shown in Appendix A1, the "model independent upper 
bound" of Waxman \& Bahcall (1999), actually, is a model dependent 
estimate based on an incorrect extrapolation of the observed flux of 
extragalactic CRs to lower energies and on 
incomplete consequences of their arbitrary choices. When properly 
corrected, this "upper bound" increases by nearly {\it two orders of 
magnitude!}  (see Appendix A1) and becomes useless.

Other  authors have tried to relate the NBR discovered with IceCube and 
the high energy gamma-ray background radiation (GBR) measured with 
Fermi-LAT, assuming that both were produced in hadronic interactions of 
high energy cosmic rays (e.g., Murase et al.~2013; Anchordoqui et 
al.~2014a and references therein). However, most of these attempts were based on the 
uncertain assumptions that the NBR is mostly extragalactic in origin and 
that the  very high energy ${\rm \gamma}$-rays from blazars and 
other extragalactic point sources are leptonic in origin. Consequently, 
the NBR was related only to the diffuse intergalactic gamma-ray 
background (IGRB), ignoring  the possibility of a much larger Galactic 
contribution and an extragalactic contribution (e.g. from blazars):

Multiwavelength observations of blazars
have shown that their spectral energy density 
usually exhibits two bumps, one at low energy 
(infrared to X-rays) and one at high energy (see, e.g., Abdo et 
al.~2010 and references therein). The origin of the high energy bump near 
TeV is still under debate  (see, e.g., Cerruti et al.~2015 and references 
therein). Moreover, current Cherenkov telescopes have identified a 
population of ultra-high-frequency peaked BL Lac objects (UHBLs), which 
exhibit exceptionally hard TeV spectra. This hard emission, as well as the 
high energy emission in gamma-ray bursts (GRBs)  challenge the  
synchrotron-self-Compton (SSC) model (e.g., Konigl 1981) of very high 
energy ${\rm \gamma}$-ray emission from blazar jets (Maraschi et al. 
1992; Bloom \& Marscher 1996) and from GRB jets (e.g., Dado \& 
Dar~2005,2009). In 
particular, the standard one zone synchrotron-self-Compton (SSC) interpretation 
of the double peaked blazer spectra (see e.g., Abdo et al.~2010 and
references therein) implies that the 
SSC peak frequency is inversely proportional to the 
synchrotron peak frequency because of the Klein Nishina suppression of
the high energy inverse Compton scattering cross section. 
Such a correlation does not seem to be present in the observational 
data. In fact, the Klein-Nishina suppression of 
inverse Compton scattering of background photons by high energy electrons
and the suppression of the population of high energy electrons 
by synchrotron radiation and inverse Compton scattering from 
background photons suggest that 
${\rm \pi^0\rightarrow 2\,\gamma}$ decay  
following inclusive production and decay of mesons in 
jets encounters with diffuse matter
may become the dominant production mechanism 
of very high energy photons emitted by  
blazars (Dar \& Shaviv 1996; Dar \& Laor~1997) and GRBs (Dado \& 
Dar~2005,2009).

If the main origin of the very high energy astronomical neutrinos and 
${\rm \gamma}$-rays and CR positrons observed near Earth is the decay of 
mesons produced in high energy CR collisions with diffuse matter in/near 
the CR sources, and/or in the ISM, and/or in the IGM, then under very 
general assumptions their fluxes, spectra and sky distributions are simply 
related.  In this paper, we reconsider such relations (e.g., Dado \& Dar 
2014) and confront them with observations rather than prejudices.  
Indeed, we show that the high energy NBR observed with IceCube and the GBR 
observed with Fermi-LAT (and the high energy Cherenkov telescopes) satisfy 
the simple relations expected from hadronic production of mesons by high 
energy cosmic rays in collisions with diffuse matter in/near source 
(rather than with diffuse radiation - see Appendix A2). They imply a non 
isotropic NBR whose sky distribution is similar to that of the Fermi-LAT 
GBR. We also show that the flux of the high energy CR positrons observed 
near Earth with PAMELA and AMS02 is that expected from secondary 
production of mesons in the local ISM by the flux of cosmic ray nucleons 
(protons and nucleons bound in atomic nuclei) observed near Earth.

\section{CR production of secondaries}
Hadronic collisions of high energy nucleons of energy ${\rm E_p}$ 
with diffuse matter produce 
${\rm \gamma}$-rays, ${\rm\nu}$'s and ${\rm e^\pm}$'s mainly through 
${\pi}$ and K decays. If the lab frame energy E of these secondary 
particles is expressed as 
${\rm x\,E_p}$, the distribution of x is independent of ${\rm E_p}$ to a 
good approximation (Feynman~1969). Consequently, 
a flux ${\rm \Phi_p\propto E_p^{-\beta}}$ of CRs in a diffuse matter 
produces through hadronic collisions 
secondary ${\rm \gamma}$-rays, ${\rm\nu}$'s, and ${\rm e^\pm}$'s
with a flux per unit volume (e.g., Dar~1983)  
\begin{equation} 
{\rm \Phi_i(E) \propto \sigma_{in}(E_p)\,n_s\,c\,\Phi_p(E_p)/ \bar{x}_i} 
\label{Eq.1} 
\end{equation} 
where ${\rm i=\gamma,\,\nu, \, or\,\, e^\pm}$, the mean value
${\rm \bar{x}_i}$ depends only on the  distribution of x
in the inclusive production  ${\rm pp\rightarrow i\,X}$ but not on E,
${\rm E_p=E/\bar{x}_i}$ is the mean energy of  cosmic ray protons 
that produce particles i with energy E, and  
${\rm \sigma_{in}\approx 30\times (E_p/GeV)^{0.058}}$ mb
is the  pp total inelastic cross  section. 

Eq.~(1) implies that CR collisions with diffuse matter produce secondary 
fluxes of ${\rm \gamma}$-rays, ${\rm\nu}$'s and ${\rm e^\pm}$'s with the 
same power-law spectrum $\rm \sim E^{-\beta+0.058}$ and different 
normalization. These fluxes are later modified by propagation effects:  
attenuation of high energy ${\rm \gamma}$- rays mainly by Compton 
scattering from free electrons and by pair production on background 
photons, oscillations of neutrinos in space that spread the neutrino flux 
over the three neutrino flavors, diffusion propagation of high energy 
${\rm e^\pm}$'s in magnetic fields, and energy loss of ${\rm e^\pm}$'s by 
synchrotron emission and inverse Compton scattering of background photons.

\section{The GBR-NBR connection}
We shall assume that the main kinds of sources of high energy cosmic rays 
(supernova explosions, merger of compact stars, and mass accretion onto 
compact stars, stellar mass black holes and massive black holes) are 
common to our Galaxy and external galaxies, that Fermi acceleration 
(Fermi~1949) and galactic confinement produce a power-law spectrum 
${\rm \Phi_p\sim E^{-\beta}}$ of galactic CRs, and that the main source of 
background ${\rm \gamma}$-rays and neutrinos above TeV is hadronic 
collisions of 
these high energy CRs with diffuse matter inside or near the CR sources, 
in the ISM of their host galaxies, and in the IGM, which produce mesons 
(mostly pions and K mesons) that decay to high energy ${\rm\gamma}$-rays, 
${\rm\nu}$'s and ${\rm e^\pm}$'s. The assumption of hadronic rather than 
leptonic production is based on the facts that inverse Compton scattering 
of background photons by high energy electrons is suppressed by the Klein 
Nishina cross section and that the population of high energy electrons is 
suppressed by synchrotron radiation and inverse Compton scattering from 
background photons. (Photoproduction has an effective threshold too high 
to explain the high energy GBR observed below TeV). The hadronic 
production hypothesis, mostly overlooked, will be confronted here with 
observations rather than with prejudices.
 
Consider first hadronic production cf high energy ${\rm \gamma}$-rays and 
neutrinos by cosmic ray collisions with diffuse matter inside or near the 
CR sources.
 
The local flux of high energy  cosmic ray nucleons (free protons and
nucleons bound in atomic nuclei that hereafter  will be denoted by p)
between several GeV and the cosmic ray knee energy  ${\rm 
E_{knee}\approx}$ 1 PeV/nucleon is well described by
\begin{equation} {\rm \Phi_p(E)\approx  C\,(E/GeV)^{-\beta}\,\, fu }
\label{Eq.2}
\end{equation} 
where  ${\rm C\approx 1.8}$, 
${\rm \beta\approx 2.70 }$  
(e.g., Olive et al.~2014),
and  ${\rm fu= 1/(GeV\, cm^2\, s\,\, sr)}$ is the flux unit.  Between
the CR 'knee' and CR ankle at E$\sim$ 3 EeV, ${\rm C\approx 114}$, 
and ${\rm \beta\approx 3.0}$ (Apel et al.~2013). We 
will ignore the small differences  
between CR protons and nucleons bound in CR nuclei (A,Z)
at the same energy per nucleon, because such nucleons contribute only 
a few percents to the total CR p flux and most of them  are bound in 
light nuclei whose inelastic cross section per nucleon is ${\rm
\sim\sigma_{pA}/A\approx \sigma_{pp}}$, i.e., roughly the same 
as that of free protons. 

CR protons escape the Galaxy by diffusion through its turbulent magnetic 
fields. For a Kolmogorov spectrum (Kolmogorov~1941) of random magnetic 
fields, their escape time satisfies ${\rm \tau_{esc}(E) \propto 
E^{-1/3}}$. In a steady state, the supply rate of high energy CR protons by the 
Galactic CR sources (s) per unit volume is equal to their escape rate from 
the Galaxy. Hence, the source spectrum of CR protons 
satisfies ${\rm \Phi_s \propto \Phi_p/\tau_{esc}\propto 
E^{-\beta_s}}$ where ${\rm \beta_s=\beta-1/3}$. 
Roughly, ${\rm\beta_s=\beta-1/3\approx 2.37}$ 
for ${\rm E<E_{knee}(p)}$, which is consistent with Fermi acceleration
modified by escape by diffusion, and ${\rm \beta_s\approx 2.67}$ for 
${\rm E_{knee}(p)<E<E_{ankle}(p)}$. 

We shall assume that above 100 GeV 
the contribution of electron bremsstrahlung and
inverse Compton scattering of background photons to the GBR, that 
decreases with increasing energy,
becomes relatively small 
compared to  the contribution from 
Galactic and extragalactic $\gamma$-ray production by 
${\rm \pi^0\rightarrow 2\gamma}$ decays.  
Neutrinos are produced mainly by 
${\rm \pi}$ and K decays.

As long as the neutrinos and ${\rm gamma}$-rays are produced
mainly through ${\rm \pi}$ decay  
by CRs with energy below the CR knee whose 
in/near source  spectral index is ${\rm \beta_j\approx 
2.7-1/3-0.06\approx 2.31}$, their fluxes satisfy 
\begin{equation} 
{\rm \Phi_\nu(E)\approx (m_{\pi^+}/2\, m_{\pi^0})^{1.31} \,\Phi_\gamma(E)
\approx 0.39 \,\Phi_\gamma(E)}\ 
\label{Eq.3} 
\end{equation} 
per ${\rm \nu}$ flavor. At higher energies 
neutrinos are produced mainly through 
${\rm \pi^{\pm}\rightarrow \mu^{\pm}\, \nu}$ and    
${\rm K^+\rightarrow \mu^+\,\nu_\mu}$ decays
followed  by ${\rm \mu^{\pm}\rightarrow e^\pm\,3\nu}$ decay,
and in ${K^+\rightarrow \pi^+\pi^0;\, 2\pi^+,\pi^-;\, 
2\pi^0\,\pi^+;\, \pi^0\,\mu^+\,\nu_\mu}$ and \\
${\rm K^0\rightarrow\pi^+\pi^-;\, \pi^0\pi^0;\, 
\pi^+\pi^-\pi^0;\, 3\pi^0}$ decays followed by the 
${\rm \pi^\pm}$ and ${\rm \mu^{\pm}}$ decays.
The ${\rm \gamma}$-rays are produced mainly by 
${\rm \pi^0\rightarrow 2\gamma}$ decays.
Well above the CR knee where ${\rm \beta_j\approx 3-1/3-0.06\approx 
2.61}$, inside/near source hadronic meson production 
by cosmic rays yields 
\begin{equation} 
{\rm \Phi_\nu(E)\approx 0.52\,\Phi_\gamma(E)}. 
\label{Eq.4} 
\end{equation} 
per ${\rm \nu}$ flavor.
Eqs.~(3),(4), cannot be tested directly with current data because 
they neglect the attenuation of high energy ${\rm \gamma}$-rays, and 
because the NBR was measured at ${\rm E_\nu >30\, TeV}$, while the 
published Fermi-LAT data on the GBR (Ackermann et al.~2012) 
and on the extragalactic gamma background (Ackermann et al.~2015)
are limited to E$<1.2$ TeV (see Figs. 1,2)
(Other recently reported measurements of the high energy GBR with 
Cherenkov telescopes such as those with 
H.E.S.S around 15 TeV (Abramowski et al.~2014) and in the ARGO-YBJ 
experiment around 1 TeV, 
(Ma~2011)  were limited to the Galactic plane  ${\rm |b|<2^o}$.)   

The attenuation of Galactic ${\rm \gamma}$-rays of energy 
below TeV is negligible, while the observed extragalactic ${\rm 
\gamma}$-ray background (EGB) is strongly absorbed above 100 GeV.
The full sky GBR measured with Fermi-LAT in the energy range
100 GeV - 2 TeV can be corrected for the attenuation of the EGB 
and used in Eq.(3) to predict 
${\rm \Phi_\nu}$ in this energy range, and then it can be 
extrapolated 
to the energy range of the NBR  detected with IceCube using
${\rm \phi_\nu(E) \propto \sigma_{in}((1+z)\,20\,E )\,
\Phi_s((1+z)\,20\,E )}$, 
which follows from  Eq.~(1) for extragalactic CR sources. In this  
relation, 1+z$\approx 2.5\pm 0.5$ is the redshift at the peak of star
formation rate (i.e., of supernova explosions and GRBs) and of the
evolution function of the emission by BL Lac objects - presumably 
the main extragalactic sources of high energy CRs.

Eq.~(1) predicts that the non attenuated EGB 
behaves like ${\rm E^{-2.31}}$ well below the CR knee. Indeed,
the EGB measured with Fermi-LAT below 820 GeV was best fit 
(Ackermann et al.~2015, Model C) with an exponential cutoff power-law, 
\begin{equation} 
{\rm \Phi_{EGB}\approx (6.42\pm 0.40)\times 10^{-7}\,
(E/GeV)^{-2.30\pm 0.02}\,e^{-E/E_c} \, fu}
\label{Eq.5}
\end{equation} 
where ${\rm E_c\approx 366\pm 100 \,  GeV}$ (${\rm \chi^2/df=6.9/23}$).
This best fit is shown in Fig.~1. 
Presumably, the power-law represents the unattenuated EGB at energies 
well below the "knee",  produced by high energy  cosmic rays in 
external galaxies. Eqs.~(3) and (5)  then yield
${\rm \phi_\nu \approx  1.03\times 10^{-11}}$ fu per ${\rm \nu}$ 
flavor at E=100 GeV  whose extrapolation to E$>50$ TeV yields an 
isotropic extragalactic [EG] neutrino flux per ${\rm \nu}$
flavor, 
\begin{equation}
{\rm E^2\,\Phi_\nu[EG]\approx (0.85 \pm .30)\times 10^{-8}
\left[E\over 100\,TeV\right]^{-0.61 \pm .05}\, GeV^2\, fu.}
\label{Eq.6}
\end{equation} 

Similarly, the Galactic [MW] contribution 
${\rm E^2\, \Phi_\gamma [MW]=E^2\,( \Phi_{GBR}-\Phi_{EGB})\approx 
2.49\times 10^{-7}\,GeV^2\, fu}$
to the GBR at E=TeV and Eq.~(3) can be used to estimate  
${\rm \Phi_\nu[MW]}$, which can be extrapolated to E$>50$ TeV, yielding
\begin{equation}  
{\rm E^2\, \Phi_\nu[MW]\approx (2.62 \pm .20)\times 10^{-8}
\left[E\over 100\,TeV\right]^{-0.61 \pm .05}\, GeV^2\, fu}
\label{Eq.7}
\end{equation}
per ${\rm \nu}$ flavor.
The predicted energy flux of the NBR (all flavors) obtained 
from the 'unattenuated GBR',  
\begin{equation}
{\rm E^2\, \Phi_{\nu}\approx (1.04 \pm 0.15)\times 10^{-7}   
\left[E\over 100\,TeV\right]^{-0.61 \pm .05}\, GeV^2\, fu},
\label{Eq.8}
\end{equation}
is compared in Fig.~2 to the all-flavors NBR measured
with IceCube (Aartsen et al.~2015).  
The separate contributions of the
Milky Way ($\approx 76\%$) and  extragalactic sources ($\approx 24\%$) 
to the NBR are shown in Fig.~3.
Assuming the all flavors NBR flux to be isotropic, its best fit 
single power-law between 25 TeV and 2.8 PeV by
the IceCube collaboration (Aartsen et al.~2015), 
${\rm E^2\, \Phi_{NBR}\approx (6.69 \pm 1.20)\times 10^{-8}\,
(E/100\,TeV)^{-0.50 \pm .09}\, GeV^2\, fu}$ 
is in  rough agreement with Eq.~(8).

The sky distribution of the NBR measured with IceCube is expected to 
coincide with that of the unattenuated high energy GBR, which is roughly 
that measured by Fermi-LAT at 100 GeV. This distribution that is peaked 
sharply around the Galactic center is shown in Fig.~4. Its peak, however, 
subtends only a small solid angle: The GBR that was measured with 
Fermi-LAT near 100 GeV (Ackermann et al.~2012) suggests that only $\sim 
4.2\%$ of the neutrino events point back towards the Galactic center 
within latitudes ${\rm -8^o \leq b\leq +8^o}$ and longitudes ${\rm -80^o 
\leq l \leq +80^o}$, which cover only $\approx 0.43\%$ of the full sky. 
Also plotted are the sky distribution of the GBR at E$>$1 GeV measured 
with EGRET aboard the Compton Gamma Ray Observatory and normalized to the 
flux measured by Fermi-Lat at 100 GeV. The diffuse gamma-ray emission 
(flux and sky distribution) from the Galactic plane (${\rm 0^o<b<2^0}$) 
and from point sources measured at higher energies (e.g., Abramowski et 
al.~2014 and references therein) with buried muon detectors (e.g., 
CASA-MIA), water Cherenkov detectors (e.g., Milagro and HWAC) and 
atmospheric Cherenkov detectors (e.g., H.E.S.S., MAGIC, and VERITAS), are 
generally consistent within errors with that extrapolated from the 
Fermi-LAT GBR assuming ${\rm \Phi_\gamma (E)\propto E^{-2.30}}$ modified 
by attenuation in the Galactic and extragalactic background light.

\section{Cosmic ray positrons near Earth}  
A detailed derivation of the expected flux of high energy cosmic ray 
positrons near Earth produced in hadronic interactions of cosmic ray 
nucleons in the local ISM, and its comparison to the flux measured with 
high precision by the AMS02 collaboration (Aguilar et al.~2015) is presented 
in Dado \& Dar~2015. Here we summarize it for completeness.

In a steady state, the local flux ${\rm \Phi_{e^+}(E)}$  of 
${\rm e^+}$'s produced in the ISM  satisfies
\begin{equation}
{\rm {d\over dE}[b(E)\,\Phi_{e^+}(E)]=J_{e^+}(E)}
\label{Eq.9}
\end{equation}
where ${\rm b(E)=-dE/dt}$ is the loss rate of ${\rm e^+}$ energy
by radiation (rad) and by escape (esc) from the Galaxy by diffusion
through its turbulent magnetic fields, and 
\begin{equation}
{\rm J_{e^+}(E)\approx  K_{e^+}\,\sigma_{in}(pp)\,n_{ism}\,c\, \Phi_p}
\label{Eq.10}
\end{equation}
is the local production rate of CR positrons in the ISM whose
nucleon density in the solar neighborhood is 
${\rm n_{ism}}$.
The  solution of Eq.~(9) is
\begin{equation}
{\rm \Phi_{e^+}(E)\approx  K_{e^+}\,\sigma_{in}(pp)\,n_{ism}\,c\,  
\tau_e\,\Phi_p(E)/(\beta_j-1)}\,.
\label{Eq.11}
\end{equation}
where  ${\rm \Phi_p}$ is given by Eq.~(2),  ${\rm n_{ism}\approx 0.9\, 
cm^{-3}}$ in the solar neighborhood (e.g. Kalberla \& Dedes~2008),  
${\rm K_{e^+}\approx 7\times 
10^{-3}}$ for ${\beta_j\approx 2.7-0.06=2.64}$  
and ${\rm \tau_e=E/(dE/dt)}$ is the mean life-time 
of positrons in the ISM due
to their escape from the Galaxy by diffusion (${\rm dE/dt\sim
-E/\tau_{esc}}$) and radiative
energy losses (inverse Compton scattering of background photons and
synchrotron radiation). It  satisfies ${\rm 
1/\tau_e=1/\tau_{esc}+1/\tau_{rad}}$, i.e.,
${\rm \tau_e={\tau_{esc}\,\tau_{rad}/ (\tau_{esc}+\tau_{rad}})}\,.$

For random Galactic magnetic fields with  a Kolmogorov spectrum
\begin{equation}
{\rm \tau_{esc}\approx 7.5 \times 10^{14}\, (E/GeV)^{-1/3}\, s}
\label{Eq.12}
\end{equation}  
where the normalization has been adjusted to the value obtained
by Lipari (2014) from a leaky box model analysis of
the flux ratio  ${\rm ^{10}Be/^9Be}$ measured
with  the Cosmic Ray Isotope Spectrometer (CRIS)  in
the energy range  70-145 MeV/nucleon (Yanasak et al.~2001).

The radiative life time due to synchrotron emission in the local
ISM magnetic field with energy density  ${\rm U\approx B^2/8\pi\approx 
0.40\,eV/cm^3}$  and inverse Compton 
scattering of the local radiation background (diffuse Galactic light 
(DGL) with energy density ${\rm U\approx 0.41\,eV/cm^3}$, 
far infra red (FIR) light with ${\rm U\approx 0.40\,eV/cm^3}$, and
cosmic microwave background (CMB) with ${\rm U\approx 0.26\,eV/cm^3}$), 
was calculated in the Thomson and Klein Nishina regimes 
following the approximations introduced by  Schlickeiser \& Ruppel (2010). 
Other energy loss mechanisms that are important only at energy well below 
10 GeV (Coulomb scattering, ionization and bremsstrahlung) as well as 
threshold effects, geomagnetic shielding and solar modulation, were 
included for completeness through a best fit phenomenological depletion 
factor ${\rm D(E)=1-exp(-(E/V)^\alpha)}$, which depends on the time of the 
measurements but does not affect the behavior at E$>$10 GeV.

In Fig.~5, the flux measured with AMS02 (Aguilar et al.~2015) is compared 
to the expected local flux of CR ${\rm e^+}$'s as given 
by Eq.(11) where  ${\rm K_{e^+}\, \Phi_p(E)}\propto E^{-\beta}$  
has been replaced by ${\rm \Phi_p(E/\bar{x}_{e^+})/\bar{x}_{e^+}}$ with  
the CR proton flux measured  with AMS02 (Aguilar et al.~2015).
As can be seen from Fig.~4, the agreement is quite good. 

\section{Conclusions} We have shown that the observed fluxes of very high 
energy backgrounds of astronomical neutrinos (NBR) and gamma rays (GBR), 
measured respectively with IceCube and Fermi-LAT, satisfy a simple 
relation, which follows from a common production in high energy hadronic 
collisions of cosmic rays in/near their main Galactic and extragalactic 
sources. An additional stringent test of the common hadronic origin of the 
high energy NBR and GBR is whether the sky distribution of the NBR 
measured with IceCube is approximately that of the unattenuated GBR near 
100 GeV. Such a test, however, requires much higher statistics than 
currently available from IceCube. We have also shown that the observed 
flux of high energy cosmic ray positrons near Earth measured with AMS02 is 
in good agreement with that expected from hadronic cosmic ray production 
in the local ISM. Hence, we conclude that {\it at present} the observed 
fluxes, spectra, and sky distributions of the high energy astronomical 
neutrinos, gamma rays and cosmic ray positrons do not provide any 
compelling evidence for decay/annihilation of dark matter particles.

\section*{Appendix A: Photo-pion origin of the NBR and GBR?}

\subsection*{A.1: Upper bound on the NBR ?}

A "model-independent upper bound" \begin{equation} {\rm {E_\nu}^2 \Phi_\nu 
< 2\times 10^{-8}\, GeV^2\, fu} \label{Eq.13} \end{equation} (per neutrino 
flavor) for the energy flux of extragalactic NBR above 100 TeV was derived 
by Waxman \& Bahcall (1999). It was based on the assumption that the NBR 
is produced by the extragalactic cosmic ray protons in proton-photon 
collisions in CR sources which are optically thin.  This "model 
independent upper bound", however is a model dependent estimate fraught 
with incorrect extrapolation (see, e.g., Ahlers, et al.~2005) and 
incomplete consequences of arbitrary choices. When properly corrected, 
this "upper bound" increases nearly by {\it two orders of magnitude!} and 
becomes practically useless.

Waxman \& Bahcall (1999)  assumed  that the flux of ultra high energy 
(UHE) cosmic rays  above the CR ankle observed at Earth 
is a universal flux of extragalactic CR protons 
with an energy independent injection-rate per comoving volume 
(Waxman 1995), 
\begin{equation}
{\rm E_{CR}^2\, {d\dot{N}_{CR}\over E_{CR}}\approx 10^{44}\,erg\, 
Mpc^{-3}\, yr^{-1}}\,.
\label{Eq.14}
\end{equation}
Assuming that this injection rate   
is valid all the way down to 100 TeV,  and that each cosmic 
ray proton photo-produces 
at most a single ${\rm \pi^+}$ 
before escaping the source, 
Waxman \& Bahcall  (1999) obtained  a  "model-independent upper  bound"
\begin{equation}
{\rm {E_\nu}^2 \Phi_\nu < 2\times 10^{-8}\, GeV^2\, fu}
\label{15}
\end{equation}
(per neutrino flavor) for the energy flux of the extragalactic NBR
in the energy range  100 TeV to 100 PeV.
If the UHE cosmic rays that are observed on Earth are extragalactic in 
origin, they must be protons. This is because of the rather short lifetime 
of extragalactic UHE CR nuclei due to their photo-disintegration in 
collisions with the extragalactic background light (Stecker~1969). A 
proton-dominated composition is also indicated by the observed break near 
$(5\pm 1)\times 10^{19}$ eV in the spectrum of the UHE cosmic rays 
observed with the Hires (Abbasi et al.~2008), Pierre Auger (e.g., Abraham 
et al.~2010), and the Telescope Array observatories,
that  coincides with 
that predicted by Greisen (1966), and Zatsepin \& Kuzmin (1966) for 
UHE extragalactic CR protons. 
However, Waxman \& Bahcall (1999) also assumed a
spectrum ${\rm dN_{CR}/dE\propto E^{-2}}$ of the UHE extragalactic  
CR protons, which
extends all the way down to PeV energies,
while the observed flux of the UHE CRs below the GZK
cutoff  is well represented by (Aab et al.~2015) ${\rm \Phi_p\approx
0.2\,(E/GeV)^{-2.71}\, fu}$. This spectral behavior of UHE CR protons  
also well represents, within errors, the measured spectrum of CR
protons below the all particle CR ankle (Apel et al.~2013)
down to $\sim$100 PeV,  presumably the CR proton ankle 
(below the proton ankle, diffusion in the Galactic turbulent magnetic 
fields with a Kolmogorov spectrum changes their spectral index to  
$\approx -2.71-1/3\approx -3.04$). In a steady state, the extragalactic 
CR spectrum is identical to the  CR source spectrum. 
The observed extragalactic flux of CR protons with a  power-law 
index -2.71 (Aab et al.~2015) extrapolated down to 100 TeV yields
an energy flux  $\approx 3.5\times 10^3$ times larger 
than that of an extragalactic flux of CR protons with 
power-law index -2 postulated by Waxman \& Bahcall (1999). 

If the production of 
high energy neutrinos above 100 TeV is dominated by the decay of
${\rm \pi^+}$ from ${\rm p\,\gamma\rightarrow  
\Delta^+(1232)\rightarrow n\,\pi^+}$, then the
$\pi^+$'s carries on average a
fraction ${\rm E'_\pi/M_{\Delta^+}\approx 0.22} $ of the energy of the
incident protons
where ${\rm E'_\pi}$ is the ${\rm \pi}$ energy in the ${\rm \Delta^+}$
rest frame. Each neutrino from the ${\rm \pi^+}$ decay carries  
an average energy ${\rm <\!E_\nu\!>\approx E_\pi/4}$.
Hence, production of neutrinos through photo-production of 
${\rm \Delta^+(1232)}$'s
by high energy cosmic ray protons in {\it optically thin sources} 
satisfies 
\begin{equation}
{\rm \Phi_\nu \approx (1/2)\, (0.22/4)^{1.71}\,\, N_\Delta \,\Phi_p\approx
3.5\times 10^{_3}\, N_\Delta \,\Phi_p}
\label{Eq.16} 
\end{equation} 
per neutrino flavor (assuming complete flavor mixing by neutrino 
oscillations), where \\
${\rm N_\Delta =\sigma_{p\gamma}\, N_\gamma}\ll 1$ is 
the mean number of  ${\rm \Delta^+}$'s  produced in collisions of a 
single UHE CR proton with a photon column density ${\rm N_\gamma}$ 
before it escapes into the  intergalactic space. 
Since the  power-law fluxes of CRs and 
neutrinos from the CR sources have the same power-law index, they  suffer 
the same redshift. Hence, for a constant ${\rm N_\Delta}$ per CR proton,
all the dependence 
of Eq.~(16) on cosmic evolution 
is through the dependence 
of the accumulated ${\rm \Phi_p}$ on cosmic evolution.

But, a high energy CR proton looses $\sim 22\%$ of its energy in 
the reaction  ${\rm p\,\gamma\rightarrow
\Delta^+(1232)\rightarrow n\,\pi^+}$ followed by ${\rm n\rightarrow 
p\,e^-\, \bar{\nu}_e}$.
Since the branching ratio 
for ${\rm \pi^+}$ production is $\approx 1/2$, in order to photo produce a 
single ${\rm \pi^+}$, each UHE proton 
must produce on average two ${\rm \Delta^+}$'s. 
Hence, such  protons 
emerge from their sources with a fraction $\approx (0.78)^2\approx 0.61$,
of their initial energy. For a spectral index -2.71, it implies that 
the in-source flux of such CR protons must be larger by a 
factor $\approx (1/0.61)^{1.7}\approx 2.34$ than that of the emergent flux  
in order to produce a single
${\rm \pi^+}$ per observed CR proton. Such a flux produces a neutrino 
flux larger by a factor 2.34 than the observed flux of UHE CR protons.  
Consequently, the maximal neutrino flux from  "thin sources", which were 
defined  by Waxman \& Bahcall (1999) as sources  producing a single 
${\rm \pi^+}$ per CR proton, the  
energy flux per neutrino flavor produced in-source is
\begin{equation}
{\rm E^2\Phi_\nu \approx 4.7\, <x_\nu>^{1.71}\, E^2\, \Phi_p(E)\approx
10^{-6}\, \left[{E\over 100\, TeV}\right]^{-0.71}\, GeV^2\, fu}.  
\label{Eq.17} 
\end{equation}
This estimate for "maximally thin" sources is larger by nearly {\it two 
orders of 
magnitude} than the Waxman-Bahcall "upper bound", which was based on an 
assumed power-law index -2 of the extragalctic CR protons above the CR 
ankle instead of the observed -2.71 (e.g., Aab et al.~2015 and references 
therein) and was used for extrapolating the energy flux of the 
extragalactic UHE CR protons down to 100 TeV, and which neglected the 
energy loss of CR protons in source in producing "at least a single ${\rm 
\pi^+}$" (i.e., two ${\rm \Delta^+}$'s) per CR proton.

\subsection*{A.2: Photo-pion origin of the NBR ?} 
If the production the NBR and GBR above 30 TeV  
is dominated by the decay of photo-pion produced in CR proton-photon        
collisions ${p\,\gamma\rightarrow
\Delta^+(1232)\rightarrow n\,\pi^+}$, and ${\rm \rightarrow p\,\pi^0}$,
respectively, in/near source (Mannheim~1993), 
then, for CR protons with in-source
spectral index ${\rm \beta_s= -2.7}$, 
the unattenuated NBR and GBR satisfy the relation 
\begin{equation}
{\rm \Phi_\nu \approx (1/2)(m_{\pi^+}/2\, m_{\pi^0})^{1.7}\, \Phi_\gamma
               \approx 0.16 \Phi_\gamma}\,. 
\label{Eq.18}
\end{equation}
per ${\rm \nu}$ flavor.  However, if the steepening of the CR spectrum 
at the CR knee is due to 
a transition from Kolmogorov diffusion to drift motion, 
while the  spectral index of the source spectrum ${\rm \beta_s=-2.3}$
does not change (Anchordoqui et al.~2014b),
then ${\rm \Phi_\nu  \approx 0.21 \Phi_\gamma}$ per ${\rm \nu}$ 
flavor. Hence, the extragalactic NBR
expected from the observed EGB with Fermi-LAT, 
\begin{equation}
{\rm E^2\,\Phi_\nu[EG]\approx (1.8 \pm .30)\times 10^{-9}
\left[E\over 100\,TeV\right]^{-0.61 \pm .05}\, GeV^2\, fu,}
\label{Eq.19}
\end{equation}
i.e., much smaller than that measured with IceCube.

The Galactic contribution of photo-pion production by
CR interactions in/near source to the NBR below $\sim$ 3 PeV 
is strongly suppressed by a too high effective energy threshold.
The effective  threshold energy for 
photo-pion production  in collisions of CRs with the typical 
gray body radiation field
of a mean photon energy ${\rm \epsilon_\gamma}$ is 
\begin{equation}
{\rm E_{th}\approx {m_\pi\,m_p\over 2\,\epsilon_\gamma}
\approx {6.5\times 10^{16}\over \epsilon_\gamma/eV}\, eV.}
\label{Eq.20}
\end{equation}
The mean photon energy of stellar light in the ISM and IGM is 
${\rm \epsilon_\gamma \sim 1\, eV}$. Hence, the minimal energy of 
the bulk of neutrinos from Galactic photo-pion production is smaller 
by a factor $\approx$ 20, yielding an  energy threshold ${\rm E_\nu > 3\, 
PeV}$ for Galactic 
photo-pion neutrinos produced  by Galactic cosmic rays in/near source and 
in the ISM.

\begin{figure}[]
\centering
\epsfig{file=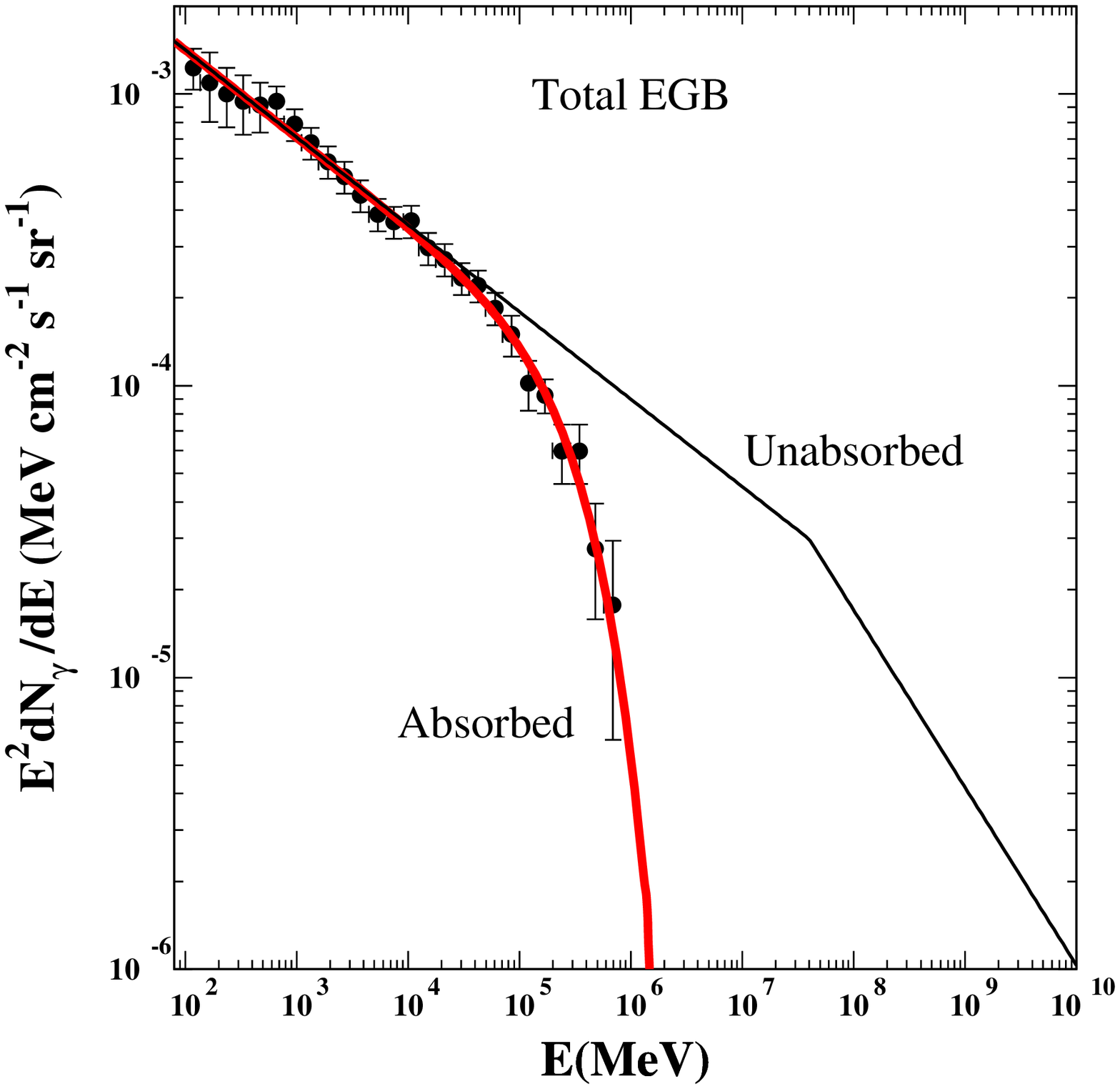,width=16.cm,height=16.cm}
\caption{The flux of the extragalactic gamma-ray 
background (EGB) as function of gamma-ray energy measured with  
Fermi-LAT (Ackermann et al.~2015) and the best 
fit exponentially cutoff power-law.
The straight line represents the unabsorbed power-law EGB.}
\label{Fig1}
\end{figure}

\begin{figure}[]
\centering
\epsfig{file=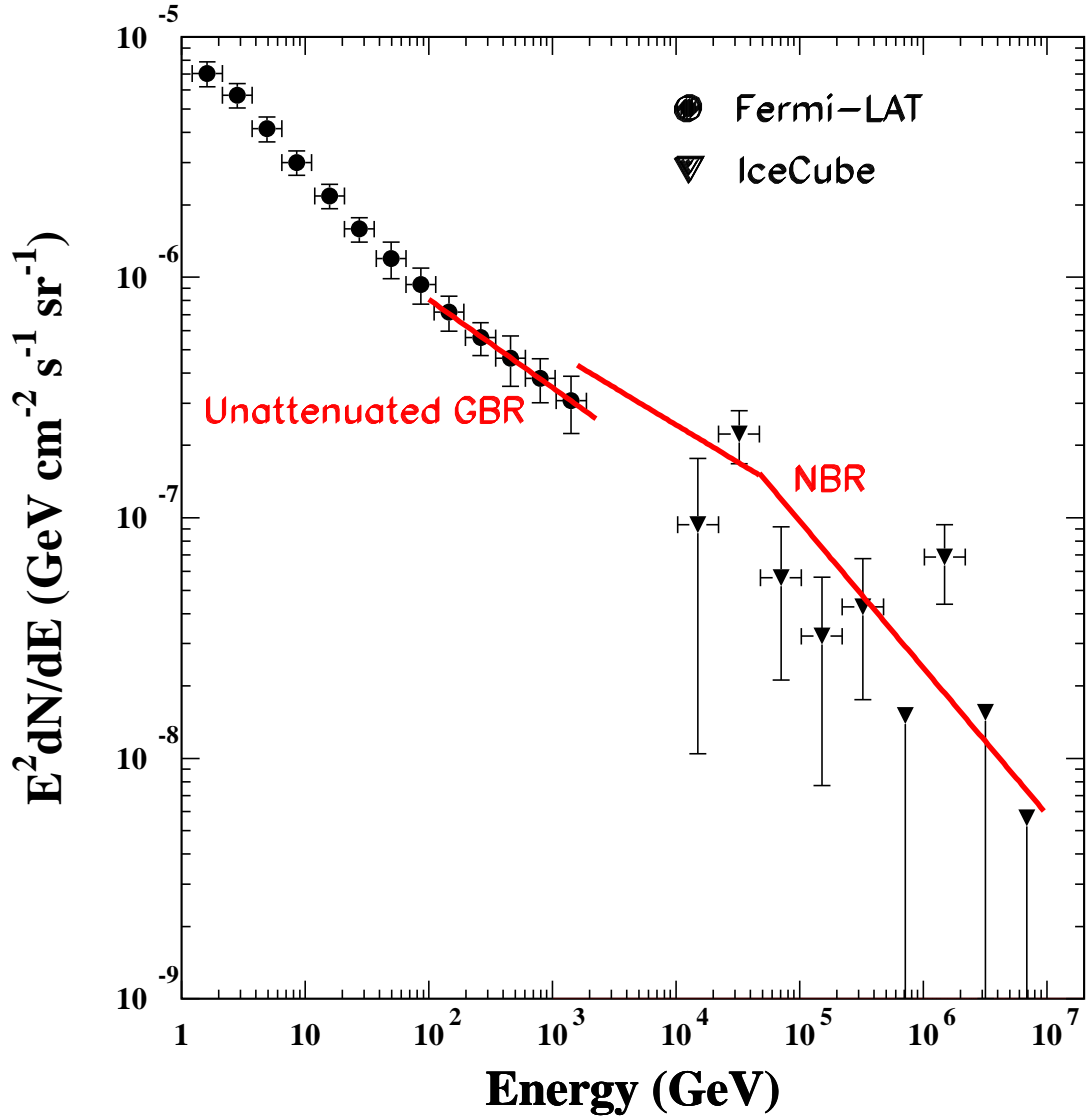,width=16.cm,height=16.cm}
\caption{Comparison between the energy flux of the NBR (all flavors)
above 20 TeV  
measured with IceCube (Aartsen et al.~2015) and that
expected from the unattenuated GBR below 2 TeV as inferred from 
the  GBR and EGB measured with Fermi-LAT (Ackermann et al. 2012,2015).
Also shown is the best fit single power-law to the unattenuated  
GBR (spectral index -2.36) in the energy range 100 GeV - 2 TeV
estimated from the Fermi-LAT measurements.}
\label{Fig2}
\end{figure}

\begin{figure}[] 
\centering
\epsfig{file=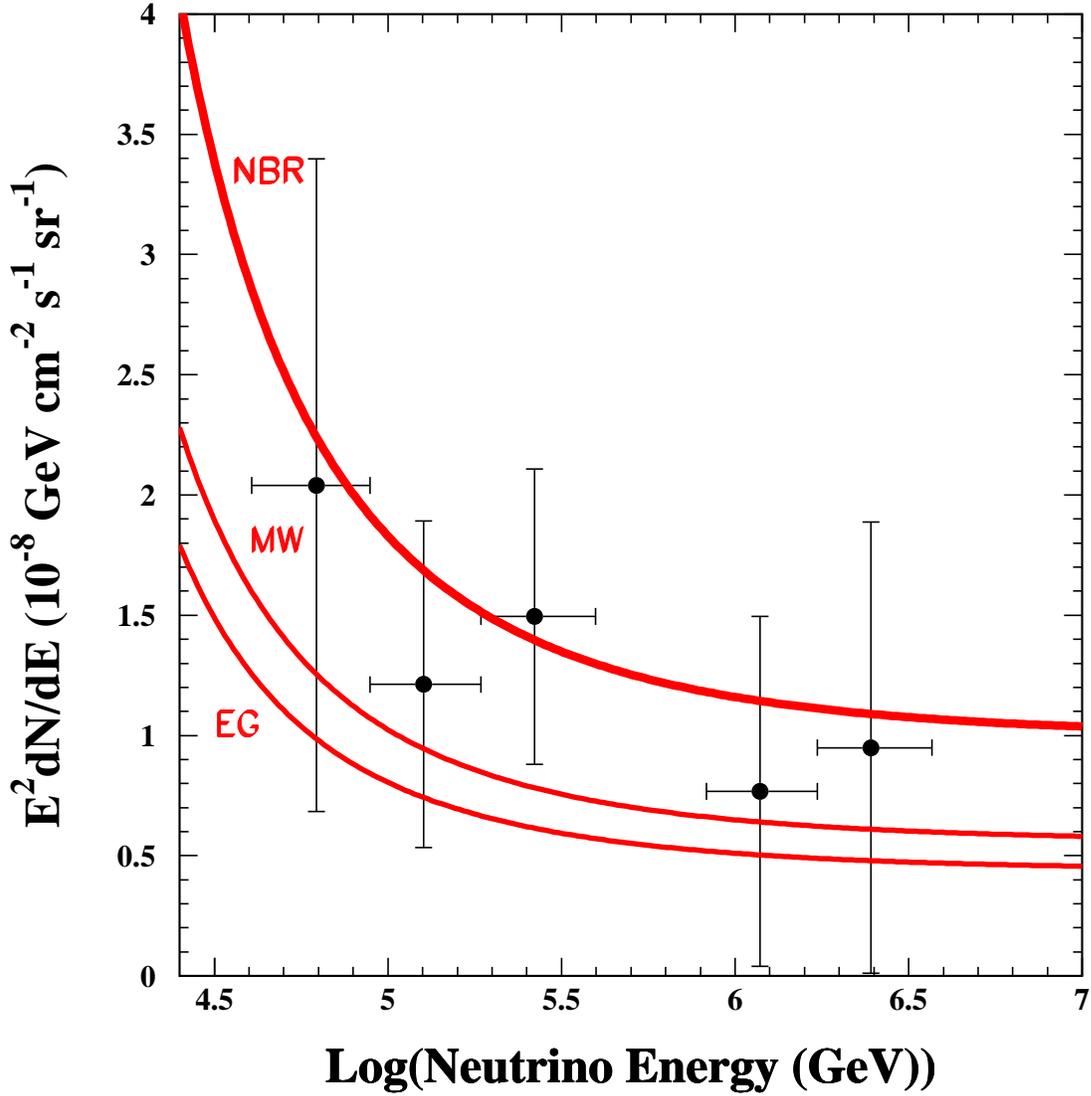,width=16.cm,height=16.cm} 
\caption{Comparison between the energy flux of the NBR (per ${\rm \nu}$
flavor) above 50 TeV, which was inferred 
from the first 3 years measurements with IceCube (Aartsen et al.~2014), 
and the flux of the NBR (per ${\rm \nu}$
flavor) expected from the GBR and  EGB, 
which were measured with Fermi-LAT (Ackermann et al. 2012,2015).
The separate contributions  of extragalactic (EG) neutrinos 
and neutrinos from our Galaxy (MW) to the NBR are also shown.} 
\label{Fig3}
\end{figure}

\begin{figure}[] 
\centering
\epsfig{file=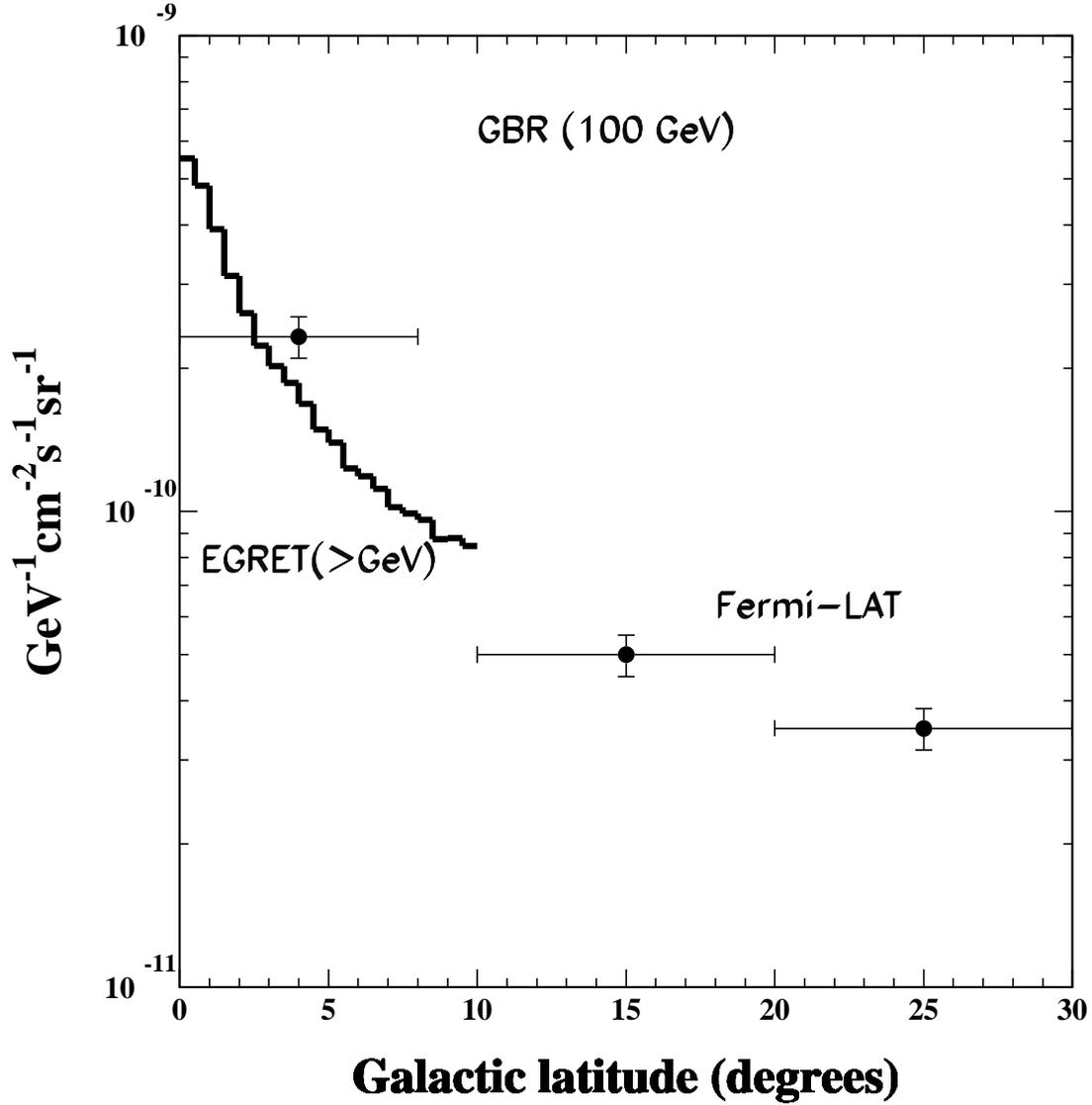,width=16.cm,height=16.cm} 
\caption{The sky distribution of the high energy GBR 
as function of Galactic latitude, 
observed with Fermi-LAT (Ackermann 2012) at 100 GeV.
Also shown is the sky distribution observed 
with EGRET aboard the Compton Gamma Ray 
Observatory at ${\rm E>1~GeV}$ 
(Pohl et al.~1997) normalized to the Fermi-LAT distribution. The NBR is 
predicted to have nearly the same sky distribution as that of the 
unattenuated high energy GBR.} 
\label{Fig4} 
\end{figure}

\begin{figure}[] 
\centering
\epsfig{file=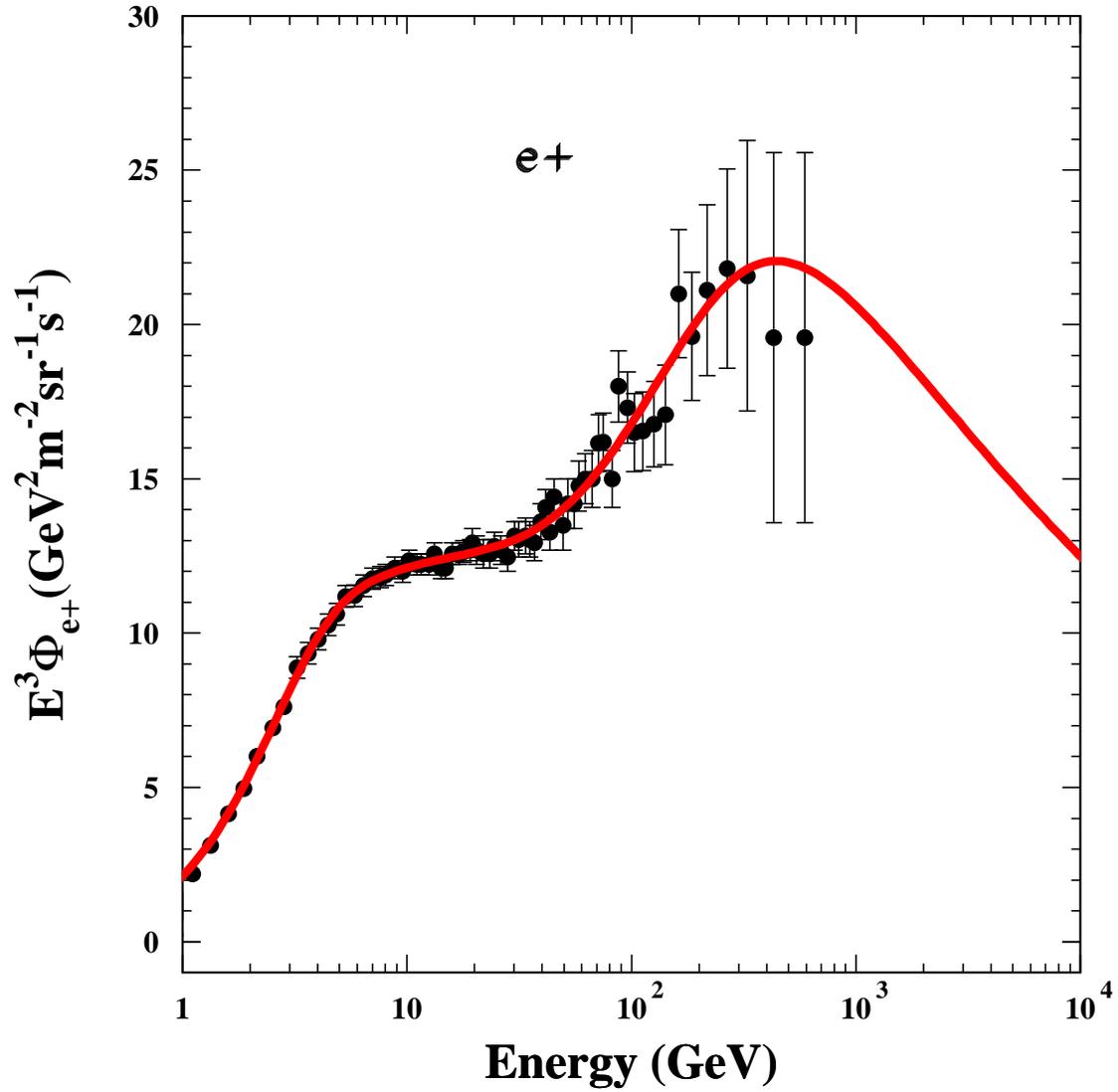,width=16.cm,height=16.cm} 
\caption{Comparison 
between the CR positron flux measured with AMS02 (Aguilar et al.~2014) and 
the flux expected from positron production in the local ISM.}
\label{Fig5} 
\end{figure}
\end{document}